\documentclass[conference,a4paper]{IEEEtran}

\IEEEoverridecommandlockouts
\IEEEpubid{\makebox[\columnwidth]{978-1-4673-6540-6/15/\$31.00 \copyright 2015 IEEE \hfill} \hspace{\columnsep}\makebox[\columnwidth]{ }}

\usepackage{cite}

\usepackage{subfig}

\ifCLASSINFOpdf
  \usepackage[pdftex]{graphicx}
\else
   \usepackage[dvips]{graphicx}
 \fi

\usepackage[cmex10]{amsmath}

\usepackage{url}
\begin{document}

\title{More is less: Connectivity in fractal regions}
\author{\IEEEauthorblockN{Carl P. Dettmann}
\IEEEauthorblockA{School of Mathematics\\
University of Bristol\\
Bristol, UK, BS8 1TW\\
Carl.Dettmann@bristol.ac.uk}
\and
\IEEEauthorblockN{Orestis Georgiou}
\IEEEauthorblockA{Toshiba Telecommunications Research Laboratory\\
32 Queens Square\\
Bristol, UK, BS1 4ND\\
Orestis.Georgiou@toshiba-trel.com}
\and
\IEEEauthorblockN{Justin P. Coon}
\IEEEauthorblockA{Department of Engineering Science\\
University of Oxford\\
Oxford UK, OX1 3PJ\\
Justin.Coon@eng.ox.ac.uk}}
\date{\today}

\maketitle

\begin{abstract}
Ad-hoc networks are often deployed in regions with complicated boundaries.
We show that if the boundary is modeled as a fractal, a network requiring line
of sight connections has the
counterintuitive property that increasing the number of nodes decreases
the full connection probability.  We characterise this decay as a stretched
exponential involving the fractal dimension of the boundary, and discuss
mitigation strategies.  Applications of this study include the analysis and
design of sensor networks operating in rugged terrain (e.g. railway
cuttings), mm-wave networks in industrial settings and vehicle-to-vehicle/vehicle-to-infrastructure
networks in urban environments.
\end{abstract}
\IEEEpeerreviewmaketitle

\section{Introduction}
Ad-hoc networks, relying on multihop connections between wireless nodes rather than direct connection to a
central router, are of growing importance in both mobile (eg vehicular) and static (eg sensor) networks~\cite{DA11}.
Advantages include scalability, flexibility and rapid deployment, as well as energy efficiency and reduced interference
associated with lower power transmission.   In many applications the spatial region where the nodes are located,
an urban mesh wifi network~\cite{VWM13},  large industrial complex~\cite{SSGG13}, or sensors monitoring a
specific geographical feature~\cite{IBSVCP10}, is highly complex, with boundary features at many length scales. 

One of the most successful and long standing models of ad hoc networks is the
random geometric graph, where nodes are placed randomly in a region and
connect pairwise if they are within a fixed range $r_0$ (although probabilistic,
``soft'' connection rules have become popular recently~\cite{CDG12b,Penrose13,HABDF09}).
There are many mathematical results, in particular, quantifying the rate at which the node
density must increase if $r_0\to 0$ and the full network is to remain
connected~\cite{Penrose97,GK99,Walters11,MA12}.

In these studies, the connection region has piecewise smooth boundaries and is typically a torus or square. Rather
more general boundaries are considered in Refs.\cite{CDG12b,CGD14} where it is shown how to accurately determine the full
connection probability $P_{fc}$ using contributions from boundary components (corners, edges, faces), any of which may
dominate depending on the node density. For example, at the highest node densities, the dominant contribution to
the outage (lack of connection) probability arises from the sharpest corners, at which the probability of isolated
nodes is greatest. Non-convex geometries have also been studied involving connecting through small
openings\cite{GDC13c,GBRDC15} and in the presence of impenetrable obstacles~\cite{DGG}. These studies highlight
the important effects of boundaries however become increasingly more difficult to analyse as the complexity of the
borders increases.
 
Here we make the leap to the extreme case where the network operation region is so complex which can be
modelled as a fractal boundary, that is, curves of dimension greater than one (see below for more precise
definitions). Fractal models have been used for natural features, for example coastlines (the ``Richardson effect'',
in which the observed length grows significantly as the measuring accuracy improves)
as popularised by Mandelbrot~\cite{Mandelbrot67},  as well as other geological features
such as mountains and rivers~\cite{PH93,Turcotte97}.  Fractal boundaries have also been used
to describe biological systems, for example trees and lungs~\cite{Losa11}.  The appearance of fractal
structures can be seen a consequence of complex dynamics as in the geological examples, or of
optimisation of geometrical parameters, for example maximum surface area at fixed volume, in biology.
Both of these are mirrored in the built environment, as fractals also provide a useful model
of urban infrastructure including land use and transport networks~\cite{Shen02}.

One particularly timely application is to sensor networks in the rail industry, to identify age-related
deterioration of tracks or emergency conditions due to overloading, natural disasters or
sabotage~\cite{HOWM14}. In this case boundaries are determined by the fractal terrain~\cite{PH93},
and a line of sight (LOS) is required for connection.  Complicated boundaries and LOS requirements are
also relevant to mm-wave networks in large industrial complexes and vehicular networks in urban
environments.  The ubiquity of fractals suggests fractals with a LOS requirement will be relevant to
many other scenarios in the future.

In random geometric graphs with smooth boundaries, the full connection probability
increases with node density at moderate and higher densities\cite{CDG12b}.  But here, it actually
{\em decreases}.  Briefly, increased node density leads to a higher probability that one or more very small sub-regions
contain a single node that cannot make an LOS connection to any others.  The purpose of this paper is to quantify
this effect, and propose steps to mitigate it.  As a first investigation of the effects of fractal boundaries, and in
order to understand the relevant effects without confounding factors, we here restrict consideration to fractal
boundaries that are exactly self-similar. 

\begin{figure}[!t]

\vspace{-90pt}

\includegraphics[width=5in]{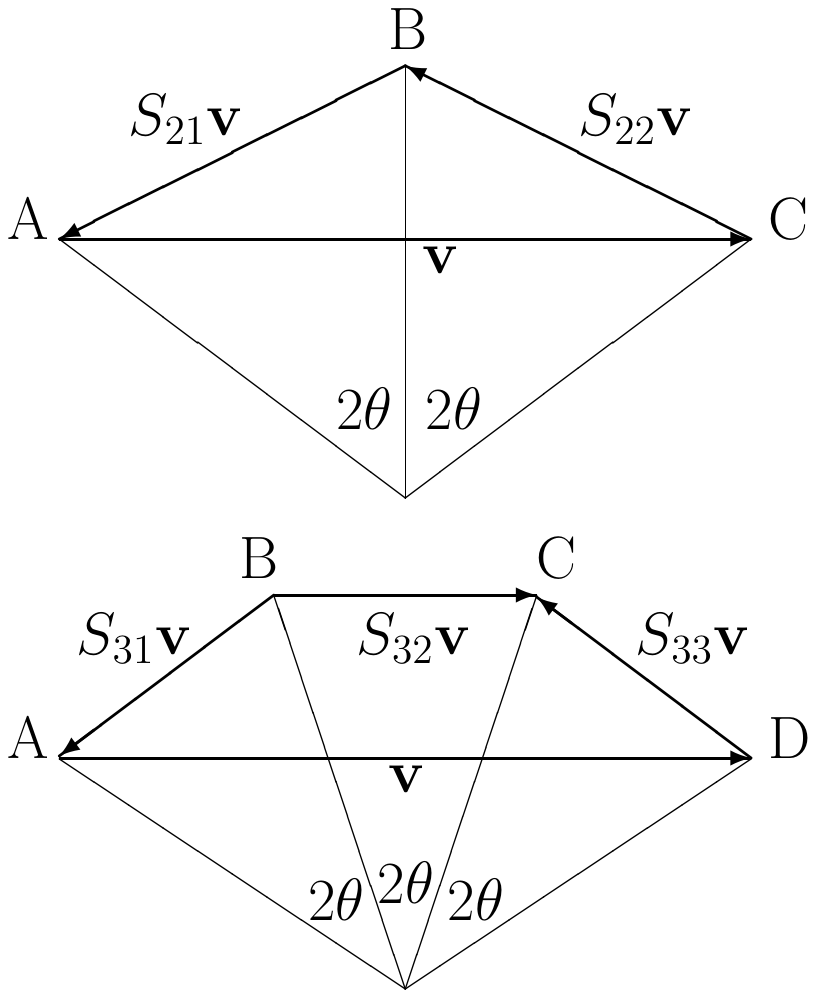}%

\vspace{-430pt}

\hspace{65pt}
\includegraphics[width=3.8in]{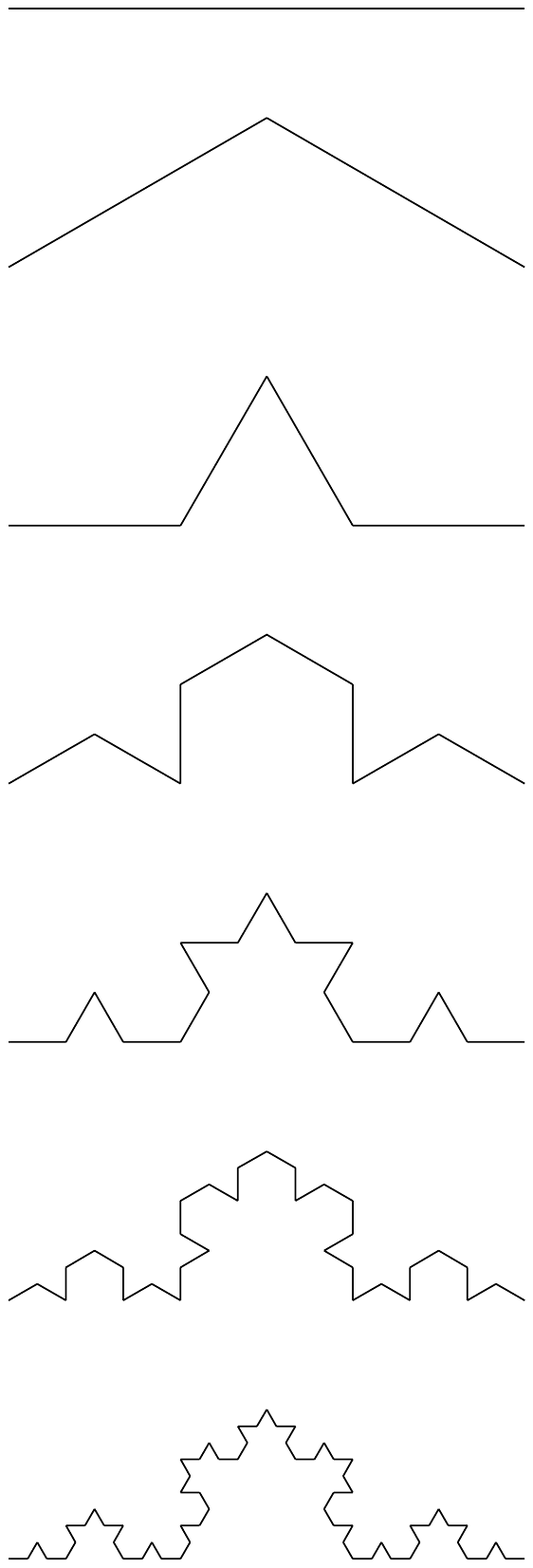}

\vspace{-20pt}

\caption{Left: Similarity transformations defined by their action on the directed line segment ${\bf v}$.
Points A, B, C and D lie on a circle, subtending angle $2\theta$ from the centre.  Right:
Construction of $\tilde{F}_2(\pi/6)$ (Koch curve) by repeated action of $S_{21}$ and $S_{22}$ on
$\bf v$.} 
\label{f:sim}
\end{figure}

\section{Fractal boundaries}
There are many variants on the definition of a fractal, and of fractal dimension~\cite{Falconer13}.
Generally there is some requirement for structures that repeat at many different length scales (though
in real systems the range of scales is finite), and that a fractal (eg Hausdorff) dimension is strictly
greater than the topological dimension (here $1$).  We do not need precise definitions of fractal
dimensions here; for completeness note that Hausdorff dimension is defined using coverings of the
set by boxes of differing sizes while box dimension is defined using boxes of equal size.  Box dimension
is always greater than or equal to Hausdorff dimension.

The examples we consider are exactly self-similar, so we can define the ``similarity dimension'' $D$, the solution of
\begin{equation}\label{e:sim}
\sum_{i=1}^n r_i^D=1
\end{equation}
Here, a fractal set $F$ is defined as the unique non-empty compact set satisfying $F=\cup_{i=1}^nS_iF$ where the $n$ similarity
transformations $S_i$ (combinations of dilations, rotations, reflections and translations) have dilation factors $0<r_i<1$.
We also assume here the ``open set condition'', that is, there is an open set $U$ so that all $S_iU$ are contained in
$U$ and are disjoint.  That is, the $S_iF$ do not overlap ``too much.''  Then, the Hausdorff and box dimensions are both
equal to $D$~\cite{Falconer13}.

We construct two families of self-similar fractals, denoted $F_2(\theta)$ with $0<\theta< \pi/4$ and $F_3(\theta)$ with
$0<\theta< \pi/6$.  Let points $A$, $B$, $C$ and $D$
lie on a circle such that the arcs $AB$, $BC$ and $CD$ each subtend angle $2\theta$ from the centre.  In two dimensions,
similarity transformations with no reflections may be uniquely defined by their action on two points.  $\tilde{F}_2$ is defined by the
transformations $S_{21}$ that maps $A\to B$ and $C\to A$ while $S_{22}$ maps $A\to C$ and $C\to B$.   $\tilde{F}_3$ is defined
by the transformations $S_{31}$ that maps $A\to B$ and $D\to A$, $S_{32}$ maps $A\to B$ and $D\to C$ and $S_{33}$ maps $A\to D$
and $D\to C$. See Fig.~\ref{f:sim}.  Then, the base interval (directed line segment $AC$ for $F_2$ or $AD$ for $F_3$) is transformed to
coincide with the interval joining $(-1,1)$ and $(1,1)$.  Finally, the fractal curve is rotated by multiples of $\pi/2$ about the origin
so that it encloses a finite area.  The union of the four copies of $\tilde{F}_2$ will be denoted $F_2$ and similarly with $F_3$.

For $\theta=0$, both ``fractals'' are actually squares of side length $2$ and boundary of dimension $1$.  $\tilde{F}_2(\pi/6)$ is the
usual Koch curve constructed by adding an equilateral triangle to the middle third of an initial interval and repeating.  $\tilde{F}_2(\pi/5)$
is the ``5-fold'' Koch curve~\cite{DF93b}.  In the limit $\theta\to\pi/4$ the curve $\tilde{F}_2$ becomes space filling and of dimension
$2$.  Similarly, in the limit $\theta\to\pi/6$, the curve $\tilde{F}_3$ approaches a Sierpinsky triangle, of dimension $\ln3/\ln 2\approx 1.585$.

For both $F_n$  examples and arbirary $\theta$, the scale factors $r_i$ are equal and given by $r=\frac{\sin\theta}{\sin n\theta}$.
Direct application of Eq.~(\ref{e:sim}) gives
\begin{eqnarray}
D(F_2(\theta))&=&\frac{\ln 2}{\ln (2\cos\theta)}\label{e:D2}\\
D(F_3(\theta))&=&\frac{\ln 3}{\ln(4\cos^2\theta-1)}\label{e:D3}
\end{eqnarray}

We can also calculate the area enclosed by each fractal (denoted $V$ for consistency with previous work~\cite{CDG12b}).  We have in both cases
$V=4+4{V}_n(\theta)$ where the first $4$ comes from the inner square and ${V}_n(\theta)$ the area enclosed between
$y=1$ (the horizontal line passing through $A$) and $\tilde{F}_n$.  In each case, we use the similarity transformation and the area
of the relevant polygon:
\begin{eqnarray}
V_2(\theta)&=&\mbox{Area}(ABC)-2r^2V_2(\theta)\\
V_3(\theta)&=&\mbox{Area}(ABCD)-r^2V_3(\theta)
\end{eqnarray}
where in the second case, there are two negative and one positive contributions of $r^2V_3(\theta)$.  Thus
\begin{eqnarray}
V_2(\theta)&=&\frac{\tan\theta}{1+2r^2}=\frac{\sin(2\theta)}{2+\cos(2\theta)}\\
V_3(\theta)&=&\frac{2\sin^3(2\theta)}{\sin^2(3\theta)}\frac{1}{1+r^2}=\frac{2\sin^3(2\theta)}{\sin^2(3\theta)+\sin^2\theta}
\end{eqnarray}
For $F_2$, $4<V<6$, while for $F_3$, $4<V<\frac{20+12\sqrt{3}}{5}\approx 8.157$.

\begin{figure}[!t]
\centerline{
\includegraphics[width=2.8in]{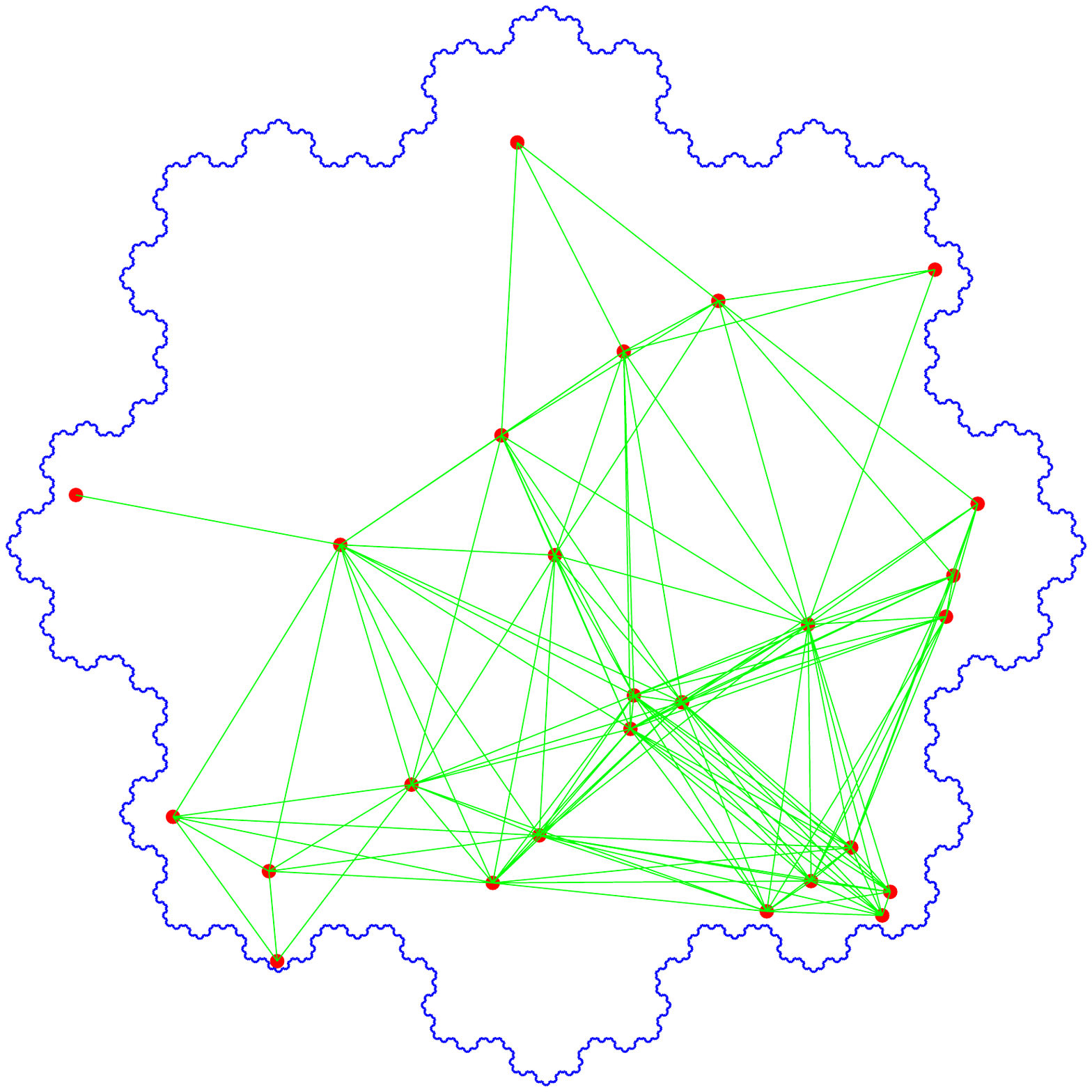}%
\hspace{-80pt}
\includegraphics[width=2.8in]{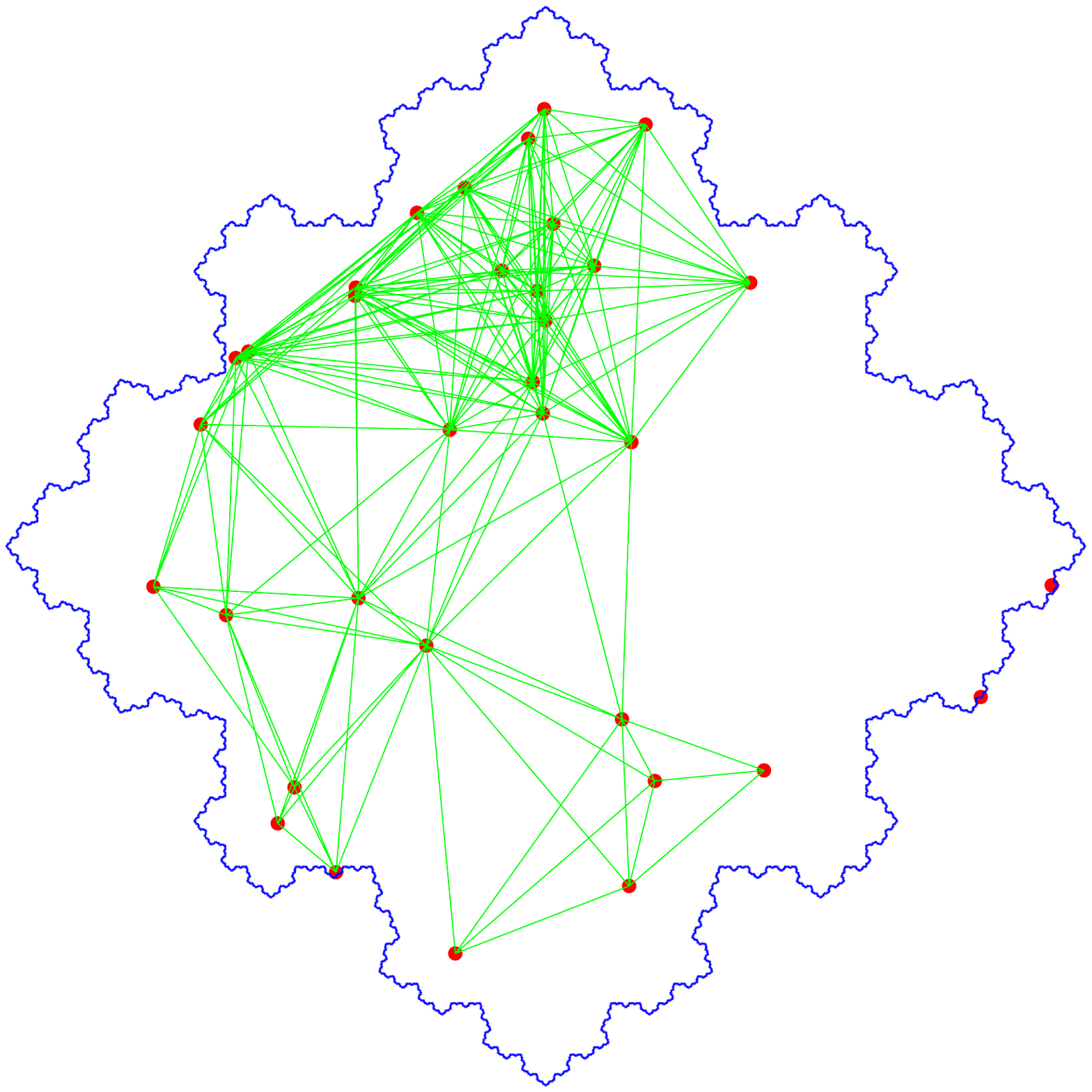}%
}

\vspace{-30pt}

\centerline{
\includegraphics[width=2.8in]{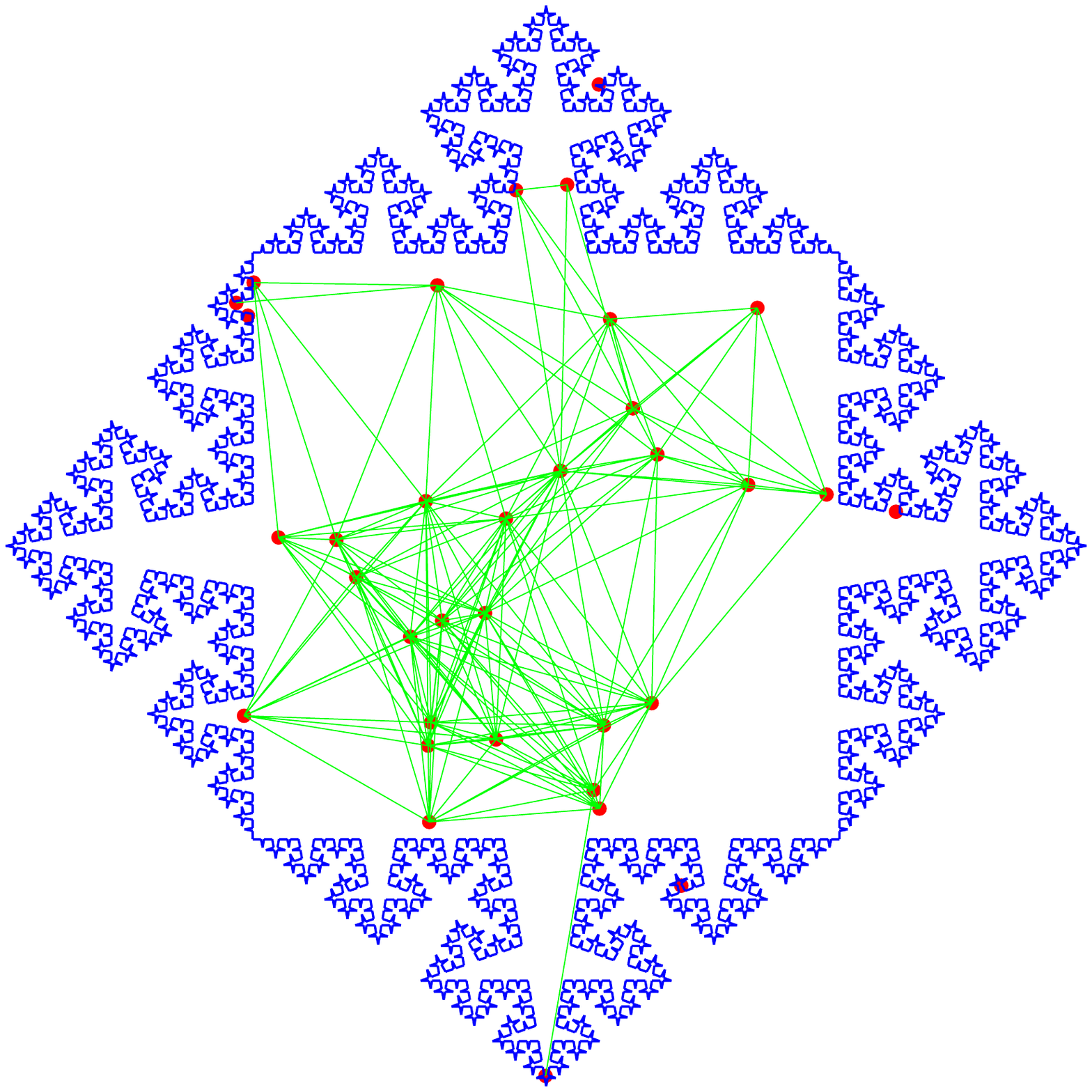}%
\hspace{-80pt}
\includegraphics[width=2.8in]{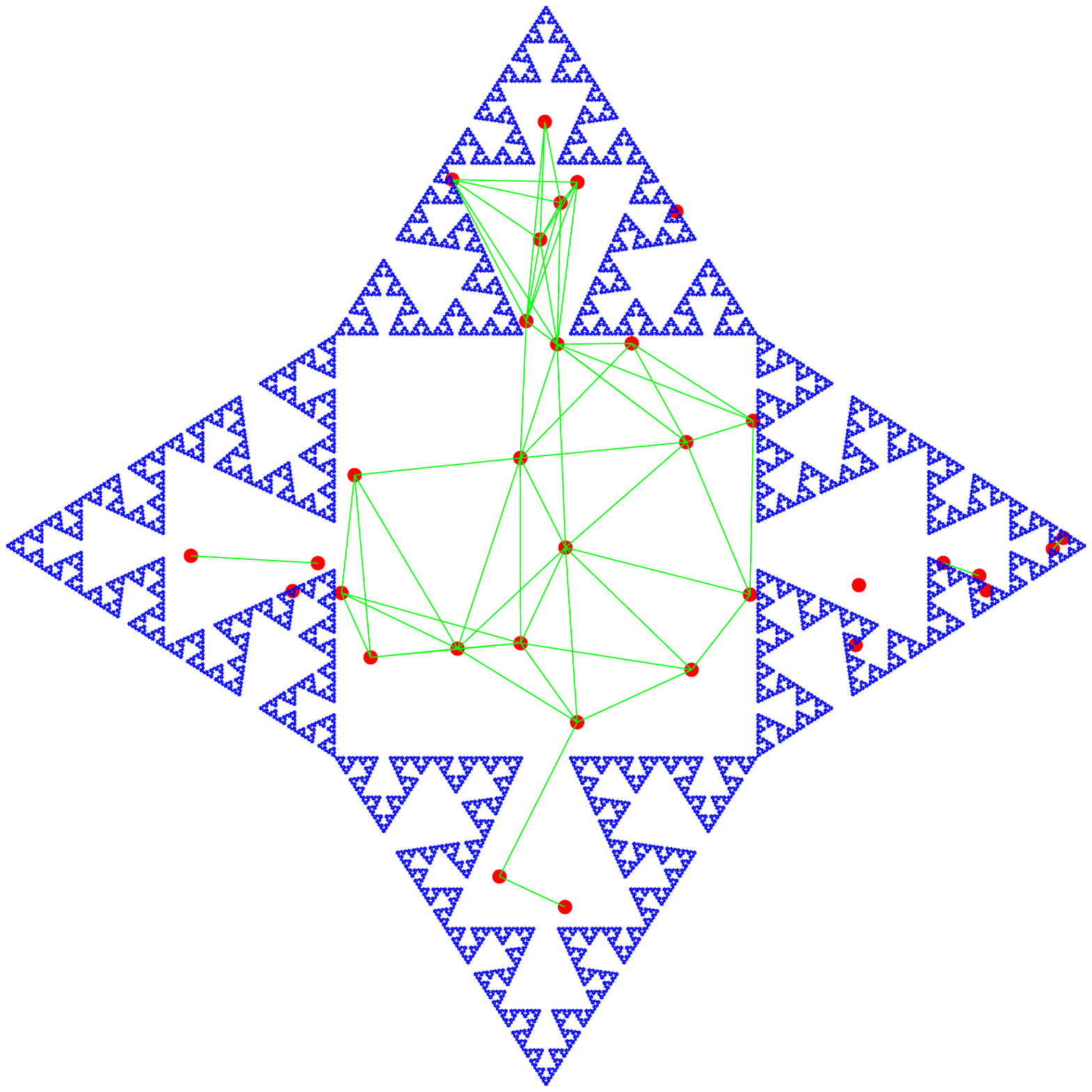}%
}

\caption{Typical networks ($\rho=5$).  Left column: $F_2(0.4)$ (top) and $F_2(0.7)$ (bottom).
Right column: $F_3(0.3)$ (top) and $F_3(0.5)$ (bottom).  Dimensions as found from Eqs.~(\protect\ref{e:D2},
\protect\ref{e:D3}) are 1.13, 1.63, 1.13, 1.50 respectively.}
\label{f:ex}
\end{figure}

\begin{figure}[!t]
\includegraphics[width=3.5in]{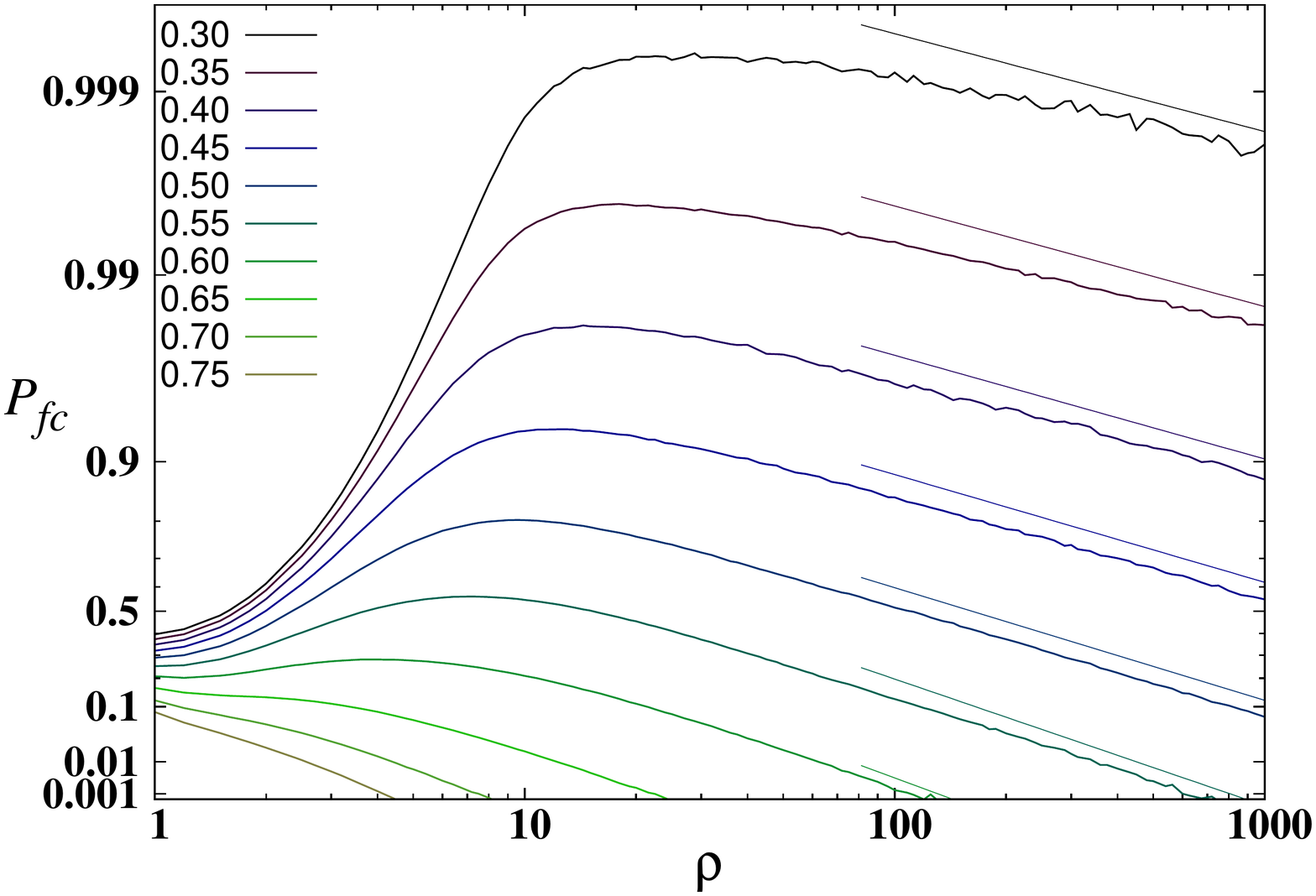}%

\vspace{-20pt}

\includegraphics[width=3.5in]{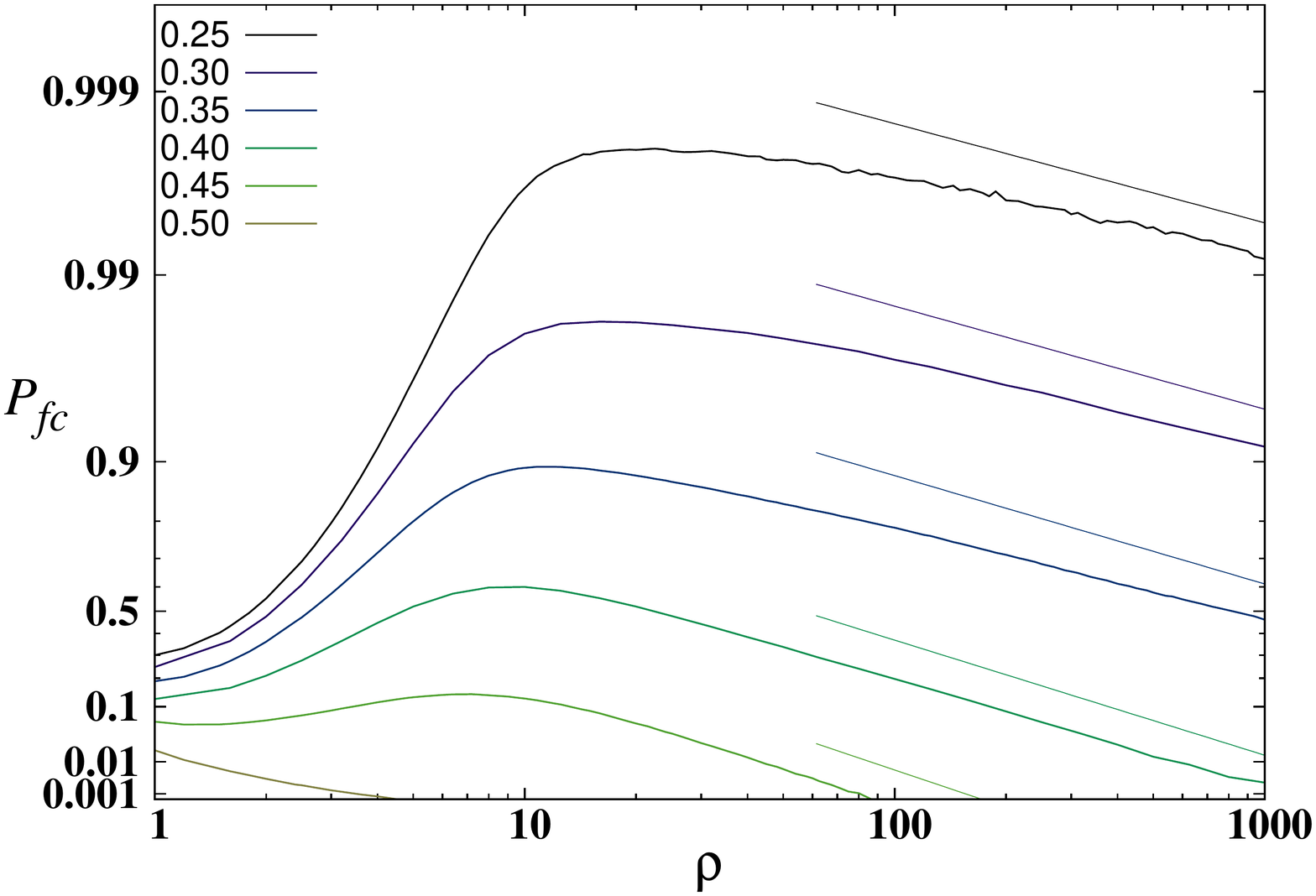}%

\caption{Full connection probability for $F_2(\theta)$ (top) and $F_3(\theta)$ (bottom) fractals as a function of node density $\rho$, for values of
$\theta$ shown in the key.  The horizontal axis is logarithmic and the vertical axis double logarithmic,
so that Eq.~\protect\ref{e:stretch} predicts straight lines of slope $D/2$ for high density, which are given for comparison.}
\label{f:Pfc}
\end{figure}

\section{Network connectivity}

Nodes are then placed according to a Poisson point process with density $\rho$ inside the fractal (hence expected number of nodes
$\rho V$), and pairwise connected if they are within range $r_0=1$ and have a line of sight connection.  Numerically, this requires two
related algorithms.  The first and simplest is to test whether a point is inside the fractal.  After locating the point in one of the four
quadrants defined by $|x|=|y|$ and rotating to the top quadrant, the point lies within the region if $y<1$ and lies outside if
$y>y_{max}=1+\tan[(n-1)\theta]$,
the highest point on the fractal.  Otherwise the appropriate inverse similarity transformations are performed until one of these conditions is
met, noting that all except $S_{32}$ switch the orientation (that is, move inside points outside and vice versa).  The other algorithm is
line of sight, tested by recursively applying the inverse similarity transformation on the whole line segment between the two nodes.
If the line segment crosses the boundaries of the transformation regions, it is split into smaller sections. The test is for whether the
interval intersects the fractal (yes if it has one endpoint with $y<1$ and one with $y>y_{max}$, no if both end points satisfy one of these
conditions), so no orientation information needs to be retained.  Typical networks constructed in this manner are shown in Fig.~\ref{f:ex}.

Here, we are concerned with the probability of full connectivity, $P_{fc}$, the fraction of possible node configurations in which all nodes
are connected in a multihop fashion, and hence ensure low latency communications throughout the network.
To determine the effect of high node density $\rho$ on connectivity, we neglect for now the possibility of nodes not closer than the finite
range $r_0$ and consider the line of sight effects near the fractal boundary.  Observe that by increasing density by a factor $r^{-d}$ leads to
the same distribution of nodes one iteration further into the fractal, where $d=2$ is the dimension of the ambient space.  There are
$n$ equivalent regions, each with a probability similar to the original of disconnecting due to an isolated node near the boundary.  Thus
we have for large node density
\begin{equation}\label{e:per}
P_{fc}(r^{-d}\rho)=P_{fc}(\rho)^n
\end{equation}
Taking logarithms we see that multiplying $\rho$ by the factor $r^{-d}$ leads to $-\ln P_{fc}$ increasing by a factor $n$.  Thus we expect
$-\ln P_{fc}$ to grow algebraically with $\rho$.  This motivates the substitution
\begin{equation}
P_{fc}=\exp[-a(\rho) \rho^{\beta}]
\end{equation}
with $a(\rho)$ and $\beta$ as yet undetermined.   Eq.~(\ref{e:per}) reduces to
\begin{equation}
a(r^{-d}\rho) r^{-d\beta}=n a(\rho)
\end{equation} 
Substituting Eq.~\ref{e:sim} and choosing $\beta=D/d$ we have simply
\begin{equation}
a(r^{-d}\rho)=a(\rho)
\end{equation}
Thus the coefficient $a$ is an unknown periodic function of $\ln\rho$ with period $d\ln r^{-1}$.
In principle we could expand it in a Fourier series, however, we expect that connection probability is a smooth function of density, so we expect
it to be dominated by its leading constant term~\cite{DF93a}.  We will henceforth treat it as constant, leading to our main result
\begin{equation}\label{e:stretch}
P_{fc}=\exp[-a\rho^{D/d}]
\end{equation}
  Thus we expect a stretched exponential decay of connection probability with
density, in contrast to smooth boundaries for which $P_{fc}\to 1$ exponentially fast.  Relevant numerical simulations are presented in
Fig.~\ref{f:Pfc}.  For comparison, straight lines corresponding to Eq.~(\ref{e:stretch}) are included.  Here, the value of $a$ was arbitrarily
chosen since the primary purpose of the illustration is to demonstrate the slope of the connectivity probability decay. Further analysis of this
parameter is not considered in the present contribution.  Notably, there is good agreement whether the ordinate $-\ln(-\ln P_{fc})$ is positive
or negative, that is, $P_{fc}$ is close to one or zero, respectively.

\section{Discussion and further applications}

We now discuss and generalise some assumptions made in the above argument.  First, some of the transformations for $F_2(\theta)$
and $F_3(\theta)$ invert the orientation of the curve, so that features that are inside the domain move outside the domain and vice versa.
For $F_2(\theta)$ both transformations invert, so we may simply apply the transformation twice, yielding the same stretched exponential.
In $F_3(\theta)$ the number of accessible outer regions at level $m$ is $2\times(3^m+(-1)^m)$ and so the ratio of accessible regions
still approaches $n=3$ in the relevant limit (high density).  Similarly, for more general fractals with differing (and generally multiplicatively
incommensurate) scale factors $r_i$, or for fractals that are only statistically self-similar, the scaling argument may apply in an average sense,
giving again Eq.~(\ref{e:stretch}).

The above analysis also applies to more complicated local network features.  For example, consider localisation of a robot swarm~\cite{BS13}.
Starting from three nodes with known locations, any node connected to at least three localisable nodes is localisable (in two dimensions),
the trilateration algorithm. Whether any node is connected to three closer to the large component is a scale invariant quantity, and so the
above argument carries through, giving the same stretched exponential with a larger value of $a$.  This applies also to more sophisticated
algorithms involving larger (but still local) network structures such as wheels~\cite{YLL09}.

As we have seen, the fractal boundary leads to a reduction in connection probability at high node densities.  At lower densities, we see from
Fig.~\ref{f:Pfc} that the connection probability increases with density (that is, it is not monotonic).  For densities that are not too small,
this is almost entirely due to isolated nodes
in the interior, as discussed by many previous authors~\cite{CDG12b,Penrose97,GK99,Walters11,MA12}.  Given the Poisson distribution, it is
easy to see that for a given node, the probability that no node lies in a circle of radius $r_0$ around it is $\exp[-\rho \pi r_0^2]$.  Such events
can be shown to be almost independent, so the expected number of isolated nodes in the interior is $\rho V\exp[-\rho \pi r_0^2]$.
However, there are many regions near the boundary for which the full area $\pi r_0^2$ is not available; a detailed analysis is likely to
be complicated and deferred to a future paper. Qualitatively we see that there is a density at which connectivity is
maximised.  Increasing the connection range $r_0$ decreases the optimum density while increasing the maximum
probability.  However, this will come at a cost to energy consumption.

Strictly speaking, the infinite density limit is singular - while our argument holds for arbitrarily large finite densities, an infinite number of
nodes placed with respect to any translationally invariant random process is connected, since the set of nodes is dense.  So, there are
no isolated nodes.  Truly infinite densities are of theoretical interest only, however, physical constraints will place a lower limit on the
relevant length scales.  At sufficiently small scales, the boundary will be smooth,
the nodes will be able to communicate through it, and size of the nodes themselves will prevent their approaching the boundary too closely.
All of these mean that at sufficiently high node densities (of the order of $\epsilon^{-d}\ln\epsilon$, where $\epsilon$ is the relevant
very small length scale) connectivity will be regained.  The number of nodes required, however is prohibitively expensive compared with
the few needed for geometries with smooth boundaries, since $\epsilon\ll r_0$.

The most resource efficient approach to regaining connectivity is likely to be the placement of gateway nodes near the entrances to the fractal
lobes.  The number of such nodes required to cover the fractal boundary is proportional to the typical length scale $\rho^{-1/d}$ to the
fractal exponent $D$, that is, the power $\rho^{-D/d}$ appearing in the stretched exponential.  This is much smaller that the total number
of nodes (on average $V\rho$).   However, this approach suffers from having to know the shape of the (often time-dependent) boundary in
great detail.

We illustrate the preceding arguments for a vehicular ad-hoc network in an urban environment.   The dimension of an urban transport network
varies significantly, but is typically around 1.25~\cite{LT04}.  Assuming an overall scale of $10^4$m and boundary roughness down to $100$m,
we need roughly $(10^4/10^2)^2=10^4$ nodes to ensure complete coverage, but only $(10^4/10^2)^{1.25}\approx  316$ gateway nodes
(in this case, static roadside units~\cite{AAAZ14}).

\section{Conclusion}
We have analysed the connectivity of dense networks confined to regions with fractal boundaries,
finding a surprising result: Increasing the density of nodes leads to lower probability of full connectivity.
We have quantified this in terms of a stretched exponential involving fractal dimension $D$ and
confirmed this numerically for two families of self-similar fractals.

It is important to know how these effects generalise to more general complex geometries,
for example self-affine or statistically self-similar fractals, and also the effect of complex
boundaries on global network properties of relevance to wireless applications, such as
centrality measures~\cite{JR13}.

We discussed a number of generalisations and amelioration strategies, however these require
a very high outlay of nodes and/or detailed knowledge of the fractal boundary.  It is likely therefore
that lack of connectivity for networks in complicated geometries will be an increasingly significant issue
in the future.

\section*{Acknowledgments}
The authors would like to thank Shinichi Baba, Jonathan Fraser, Thomas Jordan, Jed Keesling and Bill Mance
for helpful discussions and the directors of the Toshiba Telecommunications Research Laboratory
for their support. JPC and OG would like to acknowledge the support of the European Commission
partly funding the DIWINE project under Grant Agreement CNET-ICT-318177.

\bibliographystyle{IEEEtran}
\bibliography{fractal,wireless}

\end{document}